\documentclass[12pt]{article}
\usepackage{epsfig}

\def \D {\hbox{d}}
\def \Elemsimp {\mathop{\rm H}\nolimits}
\def \hh {h}

\def \sn  {\mathop{\rm sn}\nolimits}
\def \cn  {\mathop{\rm cn}\nolimits}
\def \dn  {\mathop{\rm dn}\nolimits}

\def \cd  {\mathop{\rm cd}\nolimits}
\def \nd  {\mathop{\rm nd}\nolimits}
\def \sd  {\mathop{\rm sd}\nolimits}
\def \tg  {\mathop{\rm tg}\nolimits}
\def \sech{\mathop{\rm sech}\nolimits}

\def \cotg  {\mathop{\rm cotg}\nolimits}

\def \ha  {\mathop{\rm h}\nolimits}

\def\rx{r}
\def\kx{k}
\def\kxp{{k'}}

\def \barA {\overline{A}}

\def \ee  {     \mathcal{E}}   % funny E
\def \ec  {\bar{\mathcal{E}}}  % funny E
\def \intIm           {I_{\rm m}}
\def \intIc {\overline{I_{\rm m}}}
\def \intId           {I_{\rm d}}

\def \eer  {     \mathcal{E_{\rm r}}}  % funny Er
\def \ecr  {\bar{\mathcal{E_{\rm r}}}} % funny Er
\def \intImr           {I_{\rm m,r}}
\def \intIcr {\overline{I_{\rm m,r}}}
\def \intIdr           {I_{\rm d,r}}

\def\netgain {net gain}

\def\gnl{\gamma_{\rm NL}}
\def\invuniti{\gnl \tau}

\def \mod#1{\vert #1 \vert}

\begin{document}
 
\title{Explicit solutions of the four-wave mixing model\footnote 
{
Submitted to \textit{J.~Phys.~A FTC}.
Corresponding author RC, fax +33--169088786.}} 

\author{Robert Conte\dag\ and Svetlana Bugaychuk\ddag\
{}\\
\\ \dag 
   1. LRC MESO, 
\\ \'Ecole normale sup\'erieure de Cachan (CMLA) et CEA--DAM
\\ 61, avenue du Pr\'esident Wilson, F--94235 Cachan Cedex, France.
\\ 2. Service de physique de l'\'etat condens\'e (URA 2464),
\\ CEA--Saclay, F--91191 Gif-sur-Yvette Cedex, France
\\ E-mail:  Robert.Conte@cea.fr
{}\\
\\ \ddag
Institute of Physics of the National Academy of Sciences of Ukraine
\\ 46 Prospect Nauki, Kiev-28,  UA 03028, Ukraine
\\ E-mail:  bugaich@iop.kiev.ua
{}\\
}

\maketitle

{\vglue -10.0 truemm}
{\vskip -10.0 truemm}

\begin{abstract}
The dynamical degenerate four-wave mixing 
is studied analytically in detail. 
By removing the unessential freedom, 
we first characterize this system by 
a lower-dimensional closed subsystem
of a deformed Maxwell-Bloch type,
involving only three physical variables:
the intensity pattern,
the dynamical grating amplitude,
the relative \netgain.
We then classify by the Painlev\'e test all the cases when
singlevalued solutions may exist,
according to the two essential parameters of the system:
the real relaxation time $\tau$,
the complex response constant $\gamma$.
In addition to the stationary case,
the only two integrable cases 
occur for a purely nonlocal response ($\Re(\gamma)=0$),
these are the complex unpumped Maxwell-Bloch system
and another one, which is explicitly integrated with elliptic functions.
For a generic response ($\Re(\gamma)\not=0$),
we display strong similarities
with the cubic complex Ginzburg-Landau equation.
\end{abstract}

\noindent \textit{Keywords}:
exact solutions, 
four-wave mixing,
Painlev\'e test,
singularity analysis,
solitary wave solutions.

\textit{PACS 2001} % http://publish.aps.org/PACS/01pacs.html
05.45.-a, 02.40.Xx, 42.65.-k.

\tableofcontents

% =================================================================
\section{Introduction}
\label{sec:Introduction}

The wave self-action by the
degenerate mixing in a nonlinear medium involves 
three simultaneous processes: 
the interference of waves,
the recording of the dynamical grating by an interference pattern,
and the wave diffraction by the grating. 
This process is now the basic technique of 
important practical applications
in real time holography,
including optical phase conjugation, 
holographic interferometry,
novelty filters, all-optical signal processing, etc
\cite{OSKBook,Trends87,Light2008}.

During the wave mixing,
the self-diffraction of waves
is governed by a self-consistent set of five equations
for five complex amplitudes $A_j,j=1,2,3,4$ and $\ee$,
see e.g.~\cite{OSKBook}
\begin{eqnarray}
& &
\partial_z     A_1=-i \ee     A_2,\
\partial_z \barA_2= i \ee \barA_1,\
\partial_z \barA_3=-i \ee \barA_4,\
\partial_z     A_4= i \ee     A_3,\
\label{eqFWM}
\\
& &
     \partial_t \ee = \gamma \intIm - \frac{\ee}{\tau},
\label{eqgrating_t}
\\
& &
\intIm =A_1 \barA_2 + \barA_3 A_4,
\label{eqdefIm}
\end{eqnarray}
where
(\ref{eqFWM}) is the coupled wave system for slow variable amplitudes $A_j(z,t)$
\cite{ZF}, 
(\ref{eqgrating_t}) 
is the evolution equation of the grating amplitude $\ee$
with a rhs including 
the grating gain and the grating relaxation, 
(\ref{eqdefIm}) is the relevant interference pattern of the interacting waves.
In our notation,
bar denotes complex conjugation,
$\partial$ partial derivation,
$\tau$ is a real constant.

It must be emphasized that the response constant
\begin{eqnarray}
& &
\gamma=\mod{\gamma} e^{i g}
\end{eqnarray}
is complex.
We will use the terms ``local'' and ``nonlocal'' response 
to describe the phase shift between the index grating
$\ee$ 
and the interference pattern $\intIm$.
In the case of a purely nonlocal response 
($\gamma$ purely imaginary),
an energy transfer occurs between the interacting waves, 
whereas a local response ($\gamma$ real) is characterized by 
an exchange of the phases of the waves \cite{OSKBook}.
In particular,  
the complex value of the coupling coefficient $\ee$ 
is an essential feature for the existence of soliton-like solutions. 

Apart from $t$ and $\tau$,
all variables 
are assumed dimensionless,
after normalizing the physical variables $A_j',z'$,
\begin{eqnarray}
& &
A_j = \frac{A_j'}{\sqrt{I_0}},\
z=\frac{k_0^2}{2 k_z} z',
\end{eqnarray}
where 
$k_0$ is the amplitude of the wave-vector in the free space,
$I_0$ is the total input intensity
\begin{eqnarray}
&  &
I_0 = \sum_{j=1}^{4} I_j=\hbox{constant}, I_j=\left|A_j'\right|^2.
\end{eqnarray}

We restrict here to the so-called degenerate four-wave mixing
(the four frequencies are identical),
in the transmission geometry and
in two space dimensions,
\begin{eqnarray}
& &
\vec k_j = k_{j,x} \vec e_x + k_{j,z} \vec e_z,\ j=1,2,3,4,
\\
& &
\vec k_1 - \vec k_2=\vec k_4 - \vec k_3=\vec K,
\end{eqnarray}
($\vec e_x$ and $\vec e_z$ are unit vectors,
 $\vec K$ is the grating vector).

So far, there exist two main analytic results:
\begin{description}
\item $\bullet$
for $\gamma$ purely imaginary (purely nonlocal response)
and in the stationary regime, 
a $\sech$ profile grating amplitude \cite{JBH_JOSAB};

\item $\bullet$
when the phases of each $A_j$ are independent of $z$,
a parametric representation of the five amplitudes 
also restricted to a purely imaginary $\gamma$
\cite{JBH_JOSAB,BKK,BKMPR},
\begin{eqnarray}
& & {\hskip -13.0 truemm}
\Re(\gamma)=0:\
\left\lbrace
\begin{array}{ll}
\displaystyle{
\ee=\left(\partial_z u\right) e^{\displaystyle{i \varphi_e}},\ 
\gamma=i \gnl,\ \gnl \hbox{ real},
}\\ \displaystyle{
A_1=f_{12} \sin (s_{12} (u -C_{12})) e^{\displaystyle{i\varphi_1}},\ 
A_2=f_{12} \cos (s_{12} (u -C_{12})) e^{\displaystyle{i\varphi_2}},\ 
}\\ \displaystyle{ 
A_4=-f_{43} \sin (s_{43} (u +C_{43})) e^{\displaystyle{i\varphi_4}},\ 
A_3= f_{43} \cos (s_{43} (u +C_{43})) e^{\displaystyle{i\varphi_3}},\ 
}\\ \displaystyle{ 
\varphi_1-\varphi_2-\varphi_e + \frac{\pi}{2} = n_{12} \pi,\ 
s_{12}=(-1)^{n_{12}},\
}\\ \displaystyle{
\varphi_4-\varphi_3-\varphi_e + \frac{\pi}{2} = n_{43} \pi,\
s_{43}=(-1)^{n_{43}},\
}\\ \displaystyle{ 
\intIm=\frac{1}{2} e^{\displaystyle{i(\varphi_e - \pi/2)}} 
 \left( f_{12}^2 \sin 2 (u -C_{12}) 
       -f_{43}^2 \sin 2 (u +C_{43}) \right),
}\\ \displaystyle{
n_{12}, n_{43} \in \mathcal{Z},\
}
\end{array}
\right.
\label{eqFWMparamuSG} 
\end{eqnarray}
in terms of the real solution $u$ of a damped sine-Gordon equation
\cite{JBH_JOSAB,BKK,BKMPR},
\begin{eqnarray}
& &
u_{zt} + \frac{1}{\tau} u_z -K \sin(2 u + \alpha)=0,\
K e^{i \alpha} = \frac{\gnl}{2}
\left(f_{12}^2 e^{- 2 i C_{12}} - f_{43}^2 e^{ 2 i C_{43}}\right). 
\label{eqdampedSG} 
\end{eqnarray}
The representation (\ref{eqFWMparamuSG}) displays the invariance
$(1,2,3,4,\partial_z,u) \to (4,3,2,1,-\partial_z,-u)$
and depends on six arbitrary real functions of $t$ 
($f_{12}$, $f_{43}$, $C_{12}$, $C_{43}$
and the values of $\varphi_1+\varphi_2$
and $\varphi_4+\varphi_3$)
and one arbitrary real constant (the phase $\varphi_e$).
The stationary $\sech$ solution \cite{JBH_JOSAB} is then represented by \cite{BKMPR}
(see Eq.~(\ref{eqFWM335_stationary_sol}) below),
\begin{eqnarray}
& &
\tg u = e^{\displaystyle{2 k (z - z_0)}}.
\end{eqnarray}
\end{description}

In the present article we classify all cases
when the system admits solutions with a singlevalued dependence
on the initial conditions,
and, with one major exception,
we integrate all these cases.
This major exception, 
left for future work,
presents analogous difficulties to the search,
in the complex cubic Ginzburg-Landau equation (CGL3),
\begin{eqnarray}
& &
i A_t + p A_{xx} + q \mod{A}^2 A - i \gamma A =0,\
p q \gamma \not=0,\
(A,p,q) \in {\mathcal C},\
\gamma  \in {\mathcal R},
\label{eqCGL3}
\end{eqnarray}
for source \cite{BN1985}, pulse \cite{PS1977} or front \cite{NB1984}
solutions.

% ==================================================================
\section{The intrinsic four-wave mixing, a deformed Maxwell-Bloch system}
\label{sec:Dimensional_reduction}

The ten-dimensional system
(\ref{eqFWM})--(\ref{eqdefIm})
is invariant 
under any time-dependent rotation
in the space 
$\lbrace A_1,\overline{A}_2,A_4,\overline{A}_3\rbrace$
which preserves the interference pattern (\ref{eqdefIm}).
In order to remove this five-parameter unessential freedom,
let us apply repeatedly the derivation operator $\partial_z$,
starting from the interference pattern (\ref{eqdefIm}),
until a closed system has been obtained.
This process ends after two steps and results in the 
intrinsic system
\begin{eqnarray}
& &
\partial_z \intIm = - i \ee \intId,\
\partial_z \intId=- 2 i \ec \intIm + 2 i \ee \intIc,\
\partial_t \ee = \gamma \intIm - \frac{\ee}{\tau},
\label{eqFWM555} 
\end{eqnarray}
admitting the first integral
\begin{eqnarray}
& &
4 \mod{\intIm}^2 + \intId^2 = K(t),\ K \hbox{ arbitrary}.
\label{eqFI0}
\end{eqnarray}
The real field $\intId$ which is thus naturally introduced,
\begin{eqnarray}
& &
\intId=-\mod{A_1}^2+\mod{A_2}^2-\mod{A_3}^2+\mod{A_4}^2,
\label{eqdefId}
\end{eqnarray}
has a natural interpretation:
this is the relative \textit{\netgain},
therefore the four-wave mixing is characterized by
three intrinsic variables:
the intensity pattern $\intIm$,
the grating amplitude $\ee$
and the relative \netgain\ $\intId$.

In previous integration methods \cite{CFWY} for the four-wave mixing,
one would mainly look for the wave amplitudes $A_j$
from some nonlinear system.
Thanks to the existence of the above intrinsic system, 
the integration, whether analytic or numerical,
now becomes systematic and involves two steps,
\begin{enumerate}
\item
Integration of the \textit{nonlinear} intrinsic system (\ref{eqFWM555});

\item
Knowing the grating $\ee$,
integration of the two-dimensional \textit{linear} system
\begin{eqnarray}
& &
\partial_z X=-i \ee     Y,\
\partial_z Y=-i \ec     X;
\label{eqlinearXY}
\end{eqnarray}
indeed,
given two linearly independent solutions $(X,Y)=(X_1,Y_1),(X_2,Y_2)$,
the general solution of (\ref{eqFWM}) is defined in matrix form by
\begin{eqnarray}
& &
\pmatrix{A_1 \cr A_2 \cr}
=a_{12} \pmatrix{X_1 \cr Y_1 \cr} + b_{12} \pmatrix{X_2 \cr Y_2 \cr},\
\pmatrix{\barA_3 \cr \barA_4 \cr}
=a_{34} \pmatrix{X_1 \cr Y_1 \cr} + b_{34} \pmatrix{X_2 \cr Y_2 \cr},\
\end{eqnarray}
in which the eight integration constants $a_{ij},b_{ij}$,
constrained by the three relations (\ref{eqdefIm}) and (\ref{eqdefId}),
depend on five arbitrary parameters
according to relation (\ref{eqAj_constants}) in the Appendix.
 
\end{enumerate}

The above system (\ref{eqFWM555})
is very similar to another classical system of nonlinear optics,
the pumped Maxwell-Bloch system,
which is an integrable system
defined in complex form as
\cite{BZM1987} 
\begin{eqnarray}
& &
\partial_X \rho= N e,\
\partial_X \overline{\rho}= N \overline{e},\
\partial_X N=-(\rho \overline{e} + \overline{\rho} e)/2 + 4 s=0,\
\partial_T e=\rho,\
\partial_T\overline{e}=\overline{\rho},\
\label{eqMaxwellBlochComplex}
\end{eqnarray}
with $s$ a real constant (the system is ``pumped'' when $s$ is nonzero).
 
In fact, there is only one situation when the
intrinsic four-wave mixing system (\ref{eqFWM555})
and the pumped Maxwell-Bloch system (\ref{eqMaxwellBlochComplex})
can be identified.
This occurs
when, at the same time,
the four-wave mixing model is undamped
($\tau=+\infty$)
and has a purely nonlocal response ($\Re(\gamma)=0$),
while the Maxwell-Bloch system is unpumped ($s=0$).
After this identification,
\begin{eqnarray}
& & {\hskip -8.0 truemm}
\frac{1}{\tau}=0,\
\Re(\gamma)=0,\
s=0:\ 
 \frac{z}{X}
=\frac{t}{T}
=\frac{2 \mod{\gamma} \intIm}{\rho}
=\frac{2 \mod{\gamma} \intIc}{\overline{\rho}}
=\frac{  \mod{\gamma} \intId}{N}
=\frac{- 2 i \ee}{e}
=\frac{  2 i \ec}{\overline{e}},
\label{eqIntrinsic_identical_MaxwellBloch}
\end{eqnarray}
the undamped, purely nonlocal response four-wave mixing model 
admits all the solutions of the unpumped complex Maxwell-Bloch system.

The undamped case (relaxation time $\tau=+\infty$) 
physically means the recording of a permanent grating. 
In optics that can be, for example, the permanent 
holographic memory realized in nonlinear media.

For practical computations,
it may be advisable to eliminate $\intIm$ from the grating evolution
(\ref{eqgrating_t})
and to equivalently consider the three-dimensional fifth order closed system,
\begin{eqnarray}
& &
\left\lbrace
\begin{array}{ll}
\displaystyle{
\mod{\gamma}^2 \partial_z \intId 
- 2 i      \gamma \ee (\partial_t \ec + \ec/\tau)
+ 2 i \bar \gamma \ec (\partial_t \ee + \ee/\tau)=0,
}\\ \displaystyle{
 (\partial_z \partial_t+\frac{1}{\tau}\partial_z)\ee+i    \gamma \ee \intId=0,
}\\ \displaystyle{
4 \mod{\gamma}^{-2} \mod{\partial_t \ee + \ee/\tau}^2 + \intId^2 = K(t),\ 
K \hbox{ arbitrary}.
}
\end{array}
\right.
\label{eqFWM335}
\end{eqnarray}

The sequel will display the crucial role of the third intrinsic
variable (the relative \netgain\ $\intId$)
to perform the explicit analytic integration whenever it is possible.

% ==================================================================
\section{The stationary case: general solution}
\label{sec:Stationary}

When the amplitudes are independent of the time $t$,
the integration can be performed completely.
The intrinsic system (\ref{eqFWM555})--(\ref{eqFI0})
for $\intIm,\intId,\ee$ reduces to
\begin{eqnarray}
& & 
\left\lbrace
\begin{array}{ll}
\displaystyle{
\frac{\D}{\D z}\intIm =- i  \ee \intId,\
\frac{\D}{\D z} \intId =- 4 \mod{\gamma} \tau (\sin g) \mod{\intIm}^2,\ 
\ee=\gamma \tau \intIm,\ 
}\\ \displaystyle{
4 \mod{\intIm}^2 + \intId^2 = K,
}
\end{array}
\right.
\label{eqFWM555_stationary}
\end{eqnarray}
in which the first integral $K$ is independent of $t$,
therefore $\intId$ obeys a first order
ordinary differential equation (ODE)
of the Riccati type,
\begin{eqnarray}
& &
\frac{\D}{\D z}\intId =\mod{\gamma} \tau (\sin g) (\intId^2 -K). 
\label{eqstationaryODEId}
\end{eqnarray}
The case $\gamma$ real
is uninteresting 
for it involves no energy exchange
and the intensities
$\mod{\ee}^2, \mod{\intIm}^2,\intId$ are all constant.

For $\gamma$ nonreal,
the nonlinear intrinsic system (\ref{eqFWM555_stationary})
admits the general solution
\begin{eqnarray}
& & {\hskip -10.0truemm}
\gamma \notin \mathcal{R}:
\left\lbrace
\begin{array}{ll}
\displaystyle{
\intId=-\frac{k \tanh k z}{\mod{\gamma} \tau \sin g},\ 
\ee=\gamma \tau \intIm=\frac{e^{2 i \varphi_0}}{2 \sin g}
     \left(k \sech k z\right)^{1-i \cotg g},
}\\ \displaystyle{
K=\left(\frac{k}{\mod{\gamma} \tau \sin g}\right)^2,
}
\end{array}
\right.
\label{eqFWM335_stationary_sol}
\end{eqnarray}
in which 
$k,z_0,\varphi_0$ are constants of integration,
with $z-z_0$ written for shortness as $z$.

These bright profiles for $\mod{\ee}^2$ and $\mod{\intIm}^2$ 
extrapolate the solution of Ref.~\cite{BKMPR}
which was restricted to $\gamma$ purely imaginary.

The amplitudes are found by noticing that 
each variable $A_j \ee^{-1/2}, j=1,4$ and $A_j \ec^{-1/2}$, $j=2,3$
obeys a second order linear ODE with constant coefficients.
The result is
\def\coshm{\mathop{\rm c_-}\nolimits}
\def\coshp{\mathop{\rm c_+}\nolimits}
\def\sinhm{\mathop{\rm s_-}\nolimits}
\def\sinhp{\mathop{\rm s_+}\nolimits}
\begin{eqnarray}
& &
{\hskip -8.0truemm}
\gamma \notin \mathcal{R}:
\left\lbrace
\begin{array}{ll}
\displaystyle{
A_1=\left(k \sech k z\right)^{(1-i \cotg g)/2} e^{+i \varphi_0 -i g/2}
 (\ \ \ a_{12}\coshm+\ b_{12}\sinhm),
}\\ \displaystyle{
A_2=\left(k \sech k z\right)^{(1+i \cotg g)/2} e^{-i \varphi_0 +i g/2}
 (\ -a_{12}\sinhp+\ b_{12}\coshp),
}\\ \displaystyle{
A_3=\left(k \sech k z\right)^{(1+i \cotg g)/2} e^{-i \varphi_0 +i g/2}
 (\ \ B_{34}\coshp+A_{34}\sinhp),
}\\ \displaystyle{
A_4=\left(k \sech k z\right)^{(1-i \cotg g)/2} e^{+i \varphi_0 -i g/2}
 (-B_{34}\sinhm+A_{34}\coshm),
}\\ \displaystyle{
{\rm c_\pm}=\cosh(1\pm i\cotg g)\frac{k z}{2},\
{\rm s_\pm}=\sinh(1\pm i\cotg g)\frac{k z}{2},\
}
\end{array}
\right.
\end{eqnarray}
in which the conditions that $\barA_j$ be complex conjugate of $A_j$
requires the four complex constants $a_{12},b_{12},A_{34},B_{34}$
to be represented as
\begin{eqnarray}
& &
{\hskip -8.0truemm}
\left\lbrace
\begin{array}{ll}
\displaystyle{
a_{12}=R \cos \lambda e^{ i \alpha_{12}},\
b_{12}=R \cos \mu     e^{ i \beta _{12}},\
A_{34}=R \sin \lambda e^{-i \alpha_{34}},\
B_{34}=R \sin \mu     e^{-i \beta _{34}},\
}\\ \displaystyle{
2 R^2= \frac{k}{\mod{\gamma} \tau \sin g},\
\frac{\sin(\alpha_{34}-\beta _{34})}{\sin(\alpha_{12}-\beta _{12})}
 =-\tan \lambda \tan \mu.
}
\end{array}
\right.
\end{eqnarray}
The five additional constants of integration,
chosen to be all real, are
$\lambda,\mu$,
$\alpha_{12}+\beta _{12}$, 
$\alpha_{34}+\beta _{34}$,
and for instance
$\alpha_{12}-\beta _{12}+\alpha_{34}-\beta _{34}$.

% =================================================================
\section{Determination of the cases of singlevaluedness}
\label{sec:Cases_singlevaluedness}

In the nonstationary case,
the only existing analytic result, 
valid for a purely nonlocal response ($\Re(\gamma)=0$)
and recalled in the introduction,
is the parametric representation of the five complex amplitudes
in terms of the solution $u$ of the damped sine-Gordon equation
(\ref{eqdampedSG}).
Rather than looking for solutions of this damped sine-Gordon equation,
which would only concern the case $\Re(\gamma)=0$, 
let us investigate the question of finding
singlevalued closed form solutions,
by applying the Painlev\'e test \cite{CMBook}
in order to detect all obstacles to singlevaluedness.

% =========================================================================
\subsection{The Painlev\'e test}
\label{sectionLeading}

For the basic notation 
(singular manifold variable $\varphi$,
expansion variable $\chi$,
auxiliary functions $S,C$),
we refer to detailed lecture notes \cite{Cargese1996Conte}.

Near a noncharacteristic (i.e.~$\partial_z \partial_t \not=0$)
movable singular manifold,
as shown in our preliminary article \cite{CBScicli},
the amplitudes have the leading order,
\begin{eqnarray}
& & 
\left\lbrace
\begin{array}{ll}
\displaystyle{
A_k \sim a_k \chi^{-1 + i b},\ \barA_k \sim b_k \chi^{-1 - i b},\ 
k=1,4,\
}\\ \displaystyle{
A_k \sim a_k \chi^{-1 - i b},\ \barA_k \sim b_k \chi^{-1 + i b},\ 
k=2,3,\
}\\ \displaystyle{
\ee    \sim            q_0           \chi^{-1 + 2 i b},\
\intIm \sim            I_{\rm m,0}   \chi^{-2 + 2 i b},\
\intId \sim            I_{\rm d,0}   \chi^{-2},\
}\\ \displaystyle{
\ec    \sim            r_0           \chi^{-1 - 2 i b},\
\intIc \sim {\overline{I_{\rm m,0}}} \chi^{-2 - 2 i b},\
}
\end{array}
\right.
\label{eqFWM555_leading1}
\end{eqnarray}
in which $b$ is anyone of the two real constants defined in terms of $\gamma$ by
\begin{eqnarray}
& & 
(2 b^2 -1) \cos g+ 3 b \sin g=0,\ g=\arg \gamma.   
\label{eqFWM555_leading2}
\end{eqnarray}
The leading coefficients depend 
on the nonzero auxiliary function $C(z,t)$ 
and four arbitrary complex functions $\lambda,\mu,p_{12},p_{43}$ of $(z,t)$,
\begin{eqnarray}
& &
\left\lbrace
\begin{array}{ll}
\displaystyle{
a_1= \ \ N \lambda \ \ \ p_{12}\ \   \cosh \mu,\,\
b_2=-    N \lambda \ \ \ p_{12}^{-1} \cosh \mu,\,\
}\\ \displaystyle{
a_4= \ \ N \lambda \ \ \ p_{43}\ \   \sinh \mu,\,\
b_3= \ \ N \lambda \ \ \ p_{43}^{-1} \sinh \mu,\,\
}\\ \displaystyle{
a_2= \ \ N \lambda^{-1}  p_{12}\ \   \cosh \mu,\,\
b_1= \ \ N \lambda^{-1}  p_{12}^{-1} \cosh \mu,\,\
}\\ \displaystyle{
a_3=-    N \lambda^{-1}  p_{43}\ \   \sinh \mu,\,\
b_4= \ \ N \lambda^{-1}  p_{43}^{-1} \sinh \mu,\,\
}\\ \displaystyle{
q_0=-i (1-i b) \lambda^2,\
r_0=-i (1+i b) \lambda^{-2},\
}\\ \displaystyle{
           I_{\rm m,0}  =- N^2 \lambda^2,\
{\overline{I_{\rm m,0}}}=  N^2 \lambda^{-2},\
           I_{\rm d,0}  = -2 N^2,\
}\\ \displaystyle{
N^2=\frac{C}{\mod{\gamma}} \left((1- 2 b^2)\sin g + 3 b \cos g\right),\
C \not=0.
}
\end{array}
\right.
\label{eqLeadingCoeffSol}
\end{eqnarray}

The Fuchs indices of the linearized system only depend on the value of $b$;
for the ten-dimensional system
(\ref{eqFWM})--(\ref{eqdefIm}),
these are  \cite{CBScicli}
\begin{eqnarray}
& &
j=-1,0,0,0,0,2,2,2,\frac{5 \pm \sqrt{1-48 b^2}}{2}.
\label{eq101010indices}
\end{eqnarray}
For the intrinsic five-dimensional system (\ref{eqFWM555}),
the indices are
\begin{eqnarray}
& &
j=-1,0,\frac{5 \pm \sqrt{1-48 b^2}}{2},4,
\label{eq555indices}
\end{eqnarray}
then the linear system (\ref{eqlinearXY}) admits the Fuchs indices
\begin{eqnarray}
& &
j=0,2.
\label{eq2indices}
\end{eqnarray}
The diophantine condition that all Fuchs indices be integer
therefore only admits the solution $b=0, \Re(\gamma)=0$
corresponding to a purely nonlocal response of the medium.

In order to compute the necessary conditions for the absence 
of movable logarithms arising from the integer Fuchs indices,
one can handle equivalently either 
the ten-dimensio\-nal nonlinear system (\ref{eqFWM})--(\ref{eqdefIm}),
or the five-dimensional nonlinear system (\ref{eqFWM555})
followed by the two-dimensional linear system (\ref{eqlinearXY}).
One must distinguish $b=0$ from $b\not=0$, 
and it is useless to test 
the quadruple index $0$
(because the leading order already introduces four arbitrary functions)
and the index $4$ (because of the existence of
the singlevalued first integral
$K(t)$, Eq.~(\ref{eqFI0})).
In the generic situation $b\not=0$
no movable logarithm is detected at the triple index $2$.
In the nongeneric situation $b=0$,
for instance with the five-dimensional system (\ref{eqFWM555}),
two such necessary conditions $Q_j=0$ are generated,
at the Fuchs indices $j=2$ and $3$,
\begin{eqnarray}
& &
\Re(\gamma)=0:\ 
\left\lbrace
\begin{array}{ll}
\displaystyle{
Q_2 \equiv \tau^{-1}\left(C_t + C C_z - (2 / \tau)  C \right)=0,\ 
C \not=0,\
}\\ \displaystyle{
Q_3 \equiv \tau^{-1}\left(
- \Lambda_{tt} +(2/ \tau) \Lambda_t - 2 C \Lambda_{zt} - C^2 \Lambda_{zz}
\right)=0,\ 
\lambda=e^{i \Lambda},
}
\end{array}
\right.
\label{eqFWM555Qjcond}
\end{eqnarray}
and no additional condition arises from the Fuchs index $2$ of
the linear system (\ref{eqlinearXY}).

\textit{Remark}.
The analysis of the damped sine-Gordon equation (\ref{eqdampedSG})
only generates the condition $Q_2=0$ \cite{CBScicli},
since the condition $Q_3=0$ which involves the
phases of the complex amplitudes is then identically satisfied.

A first solution to the conditions (\ref{eqFWM555Qjcond})
is $1/\tau=0$, which identifies the unpumped complex Maxwell-Bloch system
as the purely nonlocal response, undamped limit ($\Re(\gamma)=0,1/\tau=0$)
of the four-wave mixing model.

The second solution $1/\tau\not=0$ puts restrictions on the functions
$C$ and $\Lambda$.
The condition on $C$, whose general solution is \cite{CBScicli}
\begin{eqnarray}
& &
2 z/ \tau= C +F(e^{-2 t/ \tau} C),\ F \hbox{ arbitrary function},
\label{eqQ2C}
\end{eqnarray}
restricts the expansion variable $\chi$ to only depend on the
reduced variable $\xi=\sqrt{2 z} e^{-t / \tau}$
(the $\sqrt{2}$ is pure convenience)
and therefore defines a reduction $(z,t) \to \xi$
of the PDE system to an ODE system written and studied in section
\ref{sec:Time_dependent_case}.
As to the restriction on $\Lambda$,
which only makes sense for this $\xi$ reduction,
it will be further examined in section \ref{sec:Time_dependent_case}.

% =========================================================================
\subsection{Conclusion of the test}
\label{sectionConclusionTest}

The result of the test provides the guidelines to be followed
in order to obtain explicitly singlevalued solutions of the 
four-wave mixing model.
These detailed guidelines, summarized in Table \ref{TableSolutions},
are the following. 
\begin{itemize}
\item 
In the stationary case $\partial_t=0$,
the test (not performed here) succeeds,
therefore an eight-parameter single\-valued solution 
may exist.
It has already been obtained in section \ref{sec:Stationary}.

\item 
In the nonstationary, purely nonlocal response, undamped case
($\partial_t\not=0$, $\Re(\gamma)=0$, $1/ \tau =0$),
the system is equivalent to the unpumped complex Maxwell-Bloch system
(\ref{eqMaxwellBlochComplex}),
integrable in the sense of the
inverse spectral transform \cite{AblowitzClarkson},
i.e.~it admits $N$-soliton solutions,
see section \ref{sec:Maxwell_Bloch}.

\item 
In the nonstationary, purely nonlocal response, damped case,
no singlevalued solution exists
unless 
the dependence on $(z,t)$ is through the reduced variable
$\xi=\sqrt{2 z} e^{-t / \tau}$.
Then, a singlevalued solution may exist which depends on 
ten   arbitrary parameters,
we obtain it explicitly in
section \ref{sec:Time_dependent_fullPP}.

\item 
In the nonstationary, arbitrary response case, whether damped or undamped,
which includes the generic situation of the four-wave mixing,
the structure of singularities is quite similar to that
of the cubic complex Ginzburg-Landau equation (\ref{eqCGL3})
(total differential order four, 
two irrational Fuchs indices, no movable logarithm \cite{CT1989}).
Singlevalued solutions are locally represented by two Laurent series
depending on eight (instead of ten as in the two previous cases)
arbitrary functions,
and the question to find closed form solutions \textit{a priori}
presents the same difficulty as for the CGL3 equation.

\end{itemize}

\tabcolsep=1.5truemm
\tabcolsep=0.5truemm

\begin{table}[h] % [p]
\caption[Possible singlevalued solutions.]{
Possible singlevalued solutions,
according to time dependence, response ($\gamma$) and damping ($\tau$).
The reduced variable is $\xi=\sqrt{2 z} e^{-t / \tau}$.
}
\vspace{0.2truecm}
\begin{center}
\begin{tabular}{| c | c | c | c | c | c |}
\hline 
$\partial_t$&$\Re(\gamma)$&$1/\tau$& dependence & solution& Section
\\ \hline  \hline 
$=0$     &          &          & $f(z)$   &  8-param      & \ref{sec:Stationary} 
\\ \hline  \hline 
$\not=0$ & $    =0$ & $    =0$ & $f(z,t)$ & Maxwell-Bloch & \ref{sec:Maxwell_Bloch}
\\ \hline  \hline 
$\not=0$ & $    =0$ & $\not=0$ & $f(\xi)$ & 10-param      & \ref{sec:Time_dependent_fullPP} 
\\ \hline  \hline 
$\not=0$ & $\not=0$ & $     0$ & $f(z,t)$ &               & 
\\ \hline  \hline 
$\not=0$ & $\not=0$ & $\not=0$ & $f(z,t)$ &               & 
\\ \hline  \hline 
\end{tabular}
\end{center}
\label{TableSolutions}
\end{table}

% =================================================================
\section{The unpumped Maxwell-Bloch system limit}
\label{sec:Maxwell_Bloch}

Since the pumped complex Maxwell-Bloch system (\ref{eqMaxwellBlochComplex})
admits the Lax pair
\cite{KRT}
\begin{eqnarray}
& &
\partial_X \Psi=L \Psi,\
\partial_T \Psi=M \Psi,\
\nonumber
\\
& &
L = \frac{1}{2}  \pmatrix{0 & e \cr - \overline{e} & 0 \cr}
               + f \pmatrix{1 & 0 \cr 0 & -1 \cr},\
M = \frac{1}{4 f}\pmatrix{N & - \rho \cr - \overline{\rho} & -N \cr},\ 
f^2=2 s T + \lambda^2,
\label{eqLaxMaxwellBlochComplex}
\end{eqnarray}
in which $\lambda$ is an arbitrary complex constant (the spectral parameter),
the undamped four-wave mixing model with a purely nonlocal response
then admits $N$-soliton solutions, etc,
as well as solutions in terms of the 
third Painlev\'e function \cite{WinternitzSainteAdele,MilneThesis,CM2000a}.

% ==================================================================
\section{The dynamical case, reduction $\xi=(2 z)^{1/2} e^{-t/\tau}$}
\label{sec:Time_dependent_case}

The reduction $(z,t) \to \xi=(2 z)^{1/2} e^{-t/\tau}$
(with an arbitrary origin for $z$ and $t$)
isolated by the Painlev\'e test
also exists for any value of $\gamma$
and we define it so as to preserve the definitions (\ref{eqdefIm})
and (\ref{eqdefId}),
\begin{eqnarray}
& &
\frac{1}{\tau} \not=0,\ \gamma \hbox{ arbitrary}:\
\left\lbrace
\begin{array}{ll}
\displaystyle{
\intIm(z,t)=e^{-2 t/ \tau-i \omega t} \intImr(\xi),\
\intId(z,t)=e^{-2 t/ \tau} \intIdr(\xi),\
}\\ \displaystyle{
   \ee(z,t)=(1/2) e^{-  t/ \tau-i \omega t} (2 z)^{-1/2} \eer(\xi),\
}\\ \displaystyle{
    A_j(z,t)=e^{-t/ \tau-i \omega t/2}     A_{j,{\rm r}}(\xi),\ j=1,4,
}\\ \displaystyle{
    A_j(z,t)=e^{-t/ \tau+i \omega t/2}     A_{j,{\rm r}}(\xi),\ j=2,3.
}
\end{array}
\right.
\label{eqFWMReducxi}
\end{eqnarray}
It introduces one arbitrary real parameter $\omega$.

The intrinsic system (\ref{eqFWM555})--(\ref{eqFI0})
for $\intIm,\intId,\ee$ 
and the linear system for the amplitudes $A_j$
reduce to
\begin{eqnarray}
& &
\left\lbrace
\begin{array}{ll}
\displaystyle{
\frac{\D}{\D \xi} \intImr = - i \eer \intIdr,\
\frac{\D}{\D \xi} \intIdr = 2 i (\eer \intIcr - \ecr \intImr),\
\frac{\D}{\D \xi} \eer=- \gamma\tau \intImr -\frac{i \omega \tau}{\xi} \eer,\
}\\ \displaystyle{
\frac{\D}{\D \xi}     A_{1,{\rm r}}=-i \eer     A_{2,{\rm r}},\ 
\frac{\D}{\D \xi} \barA_{2,{\rm r}}= i \eer \barA_{1,{\rm r}},\
\frac{\D}{\D \xi} \barA_{3,{\rm r}}=-i \eer \barA_{4,{\rm r}},\
\frac{\D}{\D \xi}     A_{4,{\rm r}}= i \eer     A_{3,{\rm r}},
}\\ \displaystyle{
K_0=e^{4 t/ \tau} K(t)= \intIdr^2 + 4 \left|\intImr \right|^2.
}
\end{array}
\right.
\label{eqFWMxi}
\end{eqnarray}
When compared to the travelling wave reduction 
$(z,t) \to \zeta=z-ct, c \not=0$,
\begin{eqnarray}
& & {\hskip -9.0 truemm}
\left\lbrace
\begin{array}{ll}
\displaystyle{
\intIm(z,t)=e^{-i \omega t} \intImr(\zeta),\
\intId(z,t)=                \intIdr(\zeta),\
   \ee(z,t)=e^{-i \omega t} \eer(\zeta),\
}\\ \displaystyle{
     A_j(z,t)=e^{-i \omega t/2}     A_{j,{\rm r}}(\zeta),\ j=1,4,
}\\ \displaystyle{
     A_j(z,t)=e^{+i \omega t/2}     A_{j,{\rm r}}(\zeta),\ j=2,3.
}\\ \displaystyle{    
\frac{\D}{\D \zeta} \intImr = -i \eer \intIdr,\
\frac{\D}{\D \zeta} \intIdr = 2 i (\eer \intIcr - \ecr \intImr),\
\frac{\D}{\D \zeta} \eer=-\frac{\gamma}{c} \intImr
                         -(i \omega - \frac{1}{\tau}) \frac{\eer}{c},\
}\\ \displaystyle{
\frac{\D}{\D \zeta}     A_{1,{\rm r}}=-i \eer     A_{2,{\rm r}},\ 
\frac{\D}{\D \zeta} \barA_{2,{\rm r}}= i \eer \barA_{1,{\rm r}},\
\frac{\D}{\D \zeta} \barA_{3,{\rm r}}=-i \eer \barA_{4,{\rm r}},\
\frac{\D}{\D \zeta}     A_{4,{\rm r}}= i \eer     A_{3,{\rm r}}.
}\\ \displaystyle{
K_0=K(t)= \intIdr^2 + 4 \left|\intImr \right|^2,
}
\end{array}
\right.
\label{eqFWMzmct}
\end{eqnarray}
the two reduced systems (\ref{eqFWMxi}) and (\ref{eqFWMzmct})
only differ by the evolution of the grating $\eer$.

% ==================================================================
\subsection{Dynamical case, purely nonlocal response: general solution}
\label{sec:Time_dependent_fullPP}

A direct computation of the conditions (\ref{eqFWM555Qjcond})
for both reduced ODE systems (\ref{eqFWMxi}) and (\ref{eqFWMzmct})
yields
\begin{eqnarray}
{\hskip -11.0truemm}
& &
\Re(\gamma)=0:\
\left\lbrace
\begin{array}{ll}
\displaystyle{
\hbox{reduction } (2 z)^{1/2} e^{-t/\tau}:\
Q_2 \equiv 0,\ Q_3 \equiv \omega \tau \xi^{-3},
}\\ \displaystyle{
\hbox{reduction } z-ct, c \not=0:\
Q_2 \equiv \frac{1}{c \tau^3},\
Q_3 \equiv \omega 
\left(\tau \xi^{-3} -\frac{1}{2} \xi^{-2} -\frac{1}{\tau} \xi^{-1}\right),\
}
\end{array}
\right.
\label{eqreducODEnolog}
\end{eqnarray}
and the enforcement of $Q_j=0$ makes both systems identical.
Let us integrate the system (\ref{eqFWMxi}) with $\Re(\gamma)=0, \omega=0$.

Thanks to the identity of the two systems 
(\ref{eqFWMxi}) and (\ref{eqFWMzmct})
when the conditions $Q_j=0$ are enforced,
the first integrals of the system (\ref{eqFWMxi}) for $(\intImr,\intIdr,\eer)$
can be generated systematically from the reduction $X-cT$ of the Lax pair
(\ref{eqLaxMaxwellBlochComplex}) of the unpumped Maxwell-Bloch;
this provides three first integrals, all real,
\begin{eqnarray}
& & {\hskip -8.0 truemm}
\left\lbrace
\begin{array}{ll}
\displaystyle{
K_0'= (\invuniti)^2 \left(\intIdr^2 + 4 \left|\intImr \right|^2\right),\
\gamma= i \gnl,\ \gnl \hbox{ real},\
}\\ \displaystyle{
K_1= \invuniti \left(\intIcr \eer + \intImr \ecr\right),
}\\ \displaystyle{
3 e_0= \frac{1}{2} \invuniti \intIdr - \mod{\eer}^2.
}
\end{array}
\right.
\end{eqnarray}
Therefore $\intIdr$ obeys a first order ODE\footnote{
When $\Re(\gamma)=0, \omega\not=0$,
the ODE for $\intIdr$ has second order
and is studied in \cite[Eq.~(19.6)]{BureauMIII}.
}
obtained by the elimination of $\eer$ and $\intImr$,
\begin{eqnarray}
& & 
{\hskip -8.0truemm}
{\intIdr'}^2 
+ 2 \gnl\tau \intIdr^3
-12 e_0 \intIdr^2
- 2 (\gnl\tau)^{-1} K_0' \intIdr
+4 (K_1^2 + 3 e_0 K_0') (\gnl\tau)^{-2}=0.
\label{eqodeIdr}
\end{eqnarray}
The general solution $(\intImr,\intIdr,\eer)$ 
of (\ref{eqFWMxi}.1) is singlevalued
and expressible with the classical functions 
$\wp,\zeta,\sigma$ of Weierstrass,
\begin{eqnarray}
 & & \wp'^2 = 4 \wp^3 - g_2 \wp - g_3=4 (\wp-e_1)(\wp-e_2)(\wp-e_3),\
\wp=-\zeta',\
\zeta=(\log \sigma)'.
\label{eqWeierstrass2cont}
\end{eqnarray}

With the correspondence
\begin{eqnarray}
& &
K_0'=g_2 - 12 e_0^2,\
K_1^2=-\wp'(a)^2,\
-2 e_0=\wp(a),\
\label{eqnotation}
\end{eqnarray}
the squared moduli and the gradient of their phases
are doubly periodic functions,
\begin{eqnarray}
& & {\hskip -10.0 truemm}
\left\lbrace
\begin{array}{ll}
\displaystyle{
\mod{\intImr}^2=
     \frac{{\wp'}^2(\xi)-{\wp'}^2(a)}{4 (\invuniti)^2 (-\wp(\xi)+\wp(a))},\
\intIdr=\frac{-2 \wp(\xi)-\wp(a)}{\invuniti},\
\mod{\eer}^2=-\wp(\xi)+\wp(a),\
}\\ \displaystyle{
}\\ \displaystyle{
(\arg \intImr)'= - 2 K_1
  \frac{(-2 \wp(\xi) - \wp(a))(-\wp(\xi)+\wp(a))}
       {{\wp'}^2(\xi)-{\wp'}^2(a)},\
(\arg \eer)'=-\frac{K_1}{2\left(-\wp(\xi)+\wp(a)\right)},\
}\\ \displaystyle{
e^{\displaystyle{i(\arg \intImr - \arg \eer)}}=
   \frac{K_1 - i \wp'(\xi)}{2 \invuniti \mod{\eer \intImr}},
}
\end{array}
\right.
\label{eqFWM555xisolma}
\end{eqnarray}
the five constants of integration being $e_0,g_2,g_3$ (actions),
the origin of $\xi$ and
the common origin of the phase of $\intImr$ and $\eer$ (angles).

The complex amplitudes themselves $(\intImr,\eer,A_j)$ 
are also singlevalued functions and their expression,
analogous to the complex amplitude of the traveling wave of the
nonlinear Schr\"odinger equation,
involves the $\sigma$ function of Weierstrass
and is given in Appendix \ref{AppendixA}.

An important particular case occurs for $\wp'(a)=0$,
all amplitudes then have constant phases.
It proves convenient to first write this solution in complex form,
in the symmetric notation of the Jacobi functions 
as introduced by Halphen \cite{HalphenTraite},
\begin{eqnarray}
& & 
\ha_\alpha(x)=\sqrt{\wp(x)-e_\alpha},\ \alpha=1,2,3,\ 
\lim_{x \to 0} x \ha_\alpha(x)=+1,
\label{eqdefHalphenj}
\end{eqnarray}
\begin{eqnarray}
& & {\hskip -7.0 truemm}
\left\lbrace
\begin{array}{ll}
\displaystyle{
\eer   =-             e^{ 2i \varphi_0} i \ha_\alpha(\xi),\
\ecr   =-             e^{-2i \varphi_0} i \ha_\alpha(\xi),\
}\\ \displaystyle{
\intImr=-\frac{1}{\invuniti}e^{ 2i \varphi_0} \ha_\beta(\xi) \ha_\gamma(\xi),\
\intIcr= \frac{1}{\invuniti}e^{-2i \varphi_0} \ha_\beta(\xi) \ha_\gamma(\xi),\
}\\ \displaystyle{
\intIdr= \frac{1}{\invuniti} \left(-2 \ha_\alpha^2(\xi) - 3 e_\alpha \right),\
e_0=-\frac{e_\alpha}{2},\
}\\ \displaystyle{ 
    A_{1,{\rm r}}=i_0 e^{+i \varphi_0}(\   a_{12}\ha_\beta (\xi)+\ b_{12}\ha_\gamma(\xi)),\ 
    A_{2,{\rm r}}=i_0 e^{-i \varphi_0}(\ \ a_{12}\ha_\gamma(\xi)+\ b_{12}\ha_\beta (\xi)),\ 
}\\ \displaystyle{
\barA_{3,{\rm r}}=i_0 e^{+i \varphi_0}(\   a_{34}\ha_\beta (\xi)+\ b_{34}\ha_\gamma(\xi)),\ 
\barA_{4,{\rm r}}=i_0 e^{-i \varphi_0}(\ \ a_{34}\ha_\gamma(\xi)+\ b_{34}\ha_\beta (\xi)),\ 
}\\ \displaystyle{
\barA_{1,{\rm r}}=i_0 e^{-i \varphi_0}(    A_{12}\ha_\beta (\xi)+  B_{12}\ha_\gamma(\xi)),\ 
\barA_{2,{\rm r}}=i_0 e^{+i \varphi_0}(   -A_{12}\ha_\gamma(\xi)-  B_{12}\ha_\beta (\xi)),\ 
}\\ \displaystyle{
    A_{3,{\rm r}}=i_0 e^{-i \varphi_0}(    A_{34}\ha_\beta (\xi)+  B_{34}\ha_\gamma(\xi)),\ 
    A_{4,{\rm r}}=i_0 e^{+i \varphi_0}(   -A_{34}\ha_\gamma(\xi)-  B_{34}\ha_\beta (\xi)),
}\\ \displaystyle{
i_0^2=\frac{1}{\gnl\tau},    
}
\end{array}
\right.
\label{eqFWM555xisol_Halphen}
\end{eqnarray}
with $(\alpha,\beta,\gamma)$ an arbitrary permutation of $(1,2,3)$
and the relations (\ref{eqAj_constants}) for
the eight constants in $A_{j,{\rm r}}$.

In terms of the real Jacobi functions,
the complex solution (\ref{eqFWM555xisol_Halphen}) 
defines four bounded, physically admissible solutions
(i.e.~with positive square moduli for the amplitudes),
in which the grating amplitude $\eer$ is, respectively,
a $\cn,\dn,\sd,\nd$ function
(with the usual notation $\kxp^2=1-\kx^2$),
\begin{eqnarray}
& & {\hskip -7.0 truemm}
\left\lbrace
\begin{array}{ll}
\displaystyle{
\ha_1(\xi)=i \rx \kx \cn(\rx \xi,\kx),\
\ha_2(\xi)=  \rx \kx \sn(\rx \xi,\kx),\
\ha_3(\xi)=i \rx     \dn(\rx \xi,\kx),\
% ************************************************************** cn
}\\ \displaystyle{
(\alpha,\beta,\gamma)=(1,2,3):\ K_0'=\rx^4,\ \ \ \ 6 e_0=\rx^2(1-2 \kx^2),\ 
}\\ \displaystyle{
% ************************************************************** dn
(\alpha,\beta,\gamma)=(3,2,1):\ K_0'=\rx^4 \kx^4,\ 6 e_0=\rx^2(\kx^2-2),\ 
}
\end{array}
\right.
\label{eqFWM555xisol_cn_dn}
\\
& & {\hskip -7.0 truemm}
\left\lbrace
\begin{array}{ll}
\displaystyle{
\ha_1(\xi)= i \rx \kx \kxp \sd(\rx \xi,\kx),\
\ha_2(\xi)=   \rx \kx      \cd(\rx \xi,\kx),\
\ha_3(\xi)=-i \rx     \kxp \nd(\rx \xi,\kx),\
% ************************************************************** sd
}\\ \displaystyle{
(\alpha,\beta,\gamma)=(1,2,3):\ K_0'=\rx^4,\ \ \ \ 6 e_0=\rx^2(1-2 \kx^2),\ 
% ************************************************************** nd
}\\ \displaystyle{
(\alpha,\beta,\gamma)=(3,2,1):\ K_0'=\rx^4 \kx^4,\ 6 e_0=\rx^2(\kx^2-2).
}
\end{array}
\right.
\label{eqFWM555xisol_nd_sd}
\end{eqnarray}
In these nine-parameter solutions,
$\rx,\kx$ are real,
and $\lambda_{12},\lambda_{34}$ must be taken real 
in (\ref{eqAj_constants})
to ensure that $A_j$ and $\barA_j$ are complex conjugate.

A second important case is the degeneracy from doubly periodic to
simply periodic.
The subcase $\wp'(a)\not=0$, which would correspond to the dark solitary wave
\begin{eqnarray}
& & {\hskip -7.0 truemm}
\left\lbrace
\begin{array}{ll}
\displaystyle{
\eer= i e^{ 2i \varphi_0} (k \tanh(k \xi) - i \kappa) e^{ i \kappa \xi},\
\ecr= i e^{-2i \varphi_0}  (k \tanh(k \xi) + i \kappa) e^{-i \kappa \xi},\
}\\ \displaystyle{
\intImr=-\frac{1}{\invuniti} e^{ 2i \varphi_0} 
  (k^2+\kappa^2+i k \kappa \tanh(k \xi)-k^2 \tanh^2(k\xi))e^{ i \kappa \xi},\
}\\ \displaystyle{
\intIcr= \frac{1}{\invuniti} e^{-2i \varphi_0}  
  (k^2+\kappa^2-i k \kappa \tanh(k \xi)-k^2 \tanh^2(k\xi))e^{-i \kappa \xi},\
}\\ \displaystyle{
\intIdr=\frac{1}{\invuniti} (-2 k^2 \tanh^2(k\xi) + 2 k^2 +\kappa^2),
}
\end{array}
\right.
\label{eqFWM555xisolTrigoDark}
\end{eqnarray}
is unphysical since the square modulus $\eer \ecr$ is negative.
As to the subcase $\wp'(a)=0$, it defines the bright solitary wave
obtained from the long wave limit $\kx^2=1$ in (\ref{eqFWM555xisol_cn_dn}),
\begin{eqnarray}
\ha_1(\xi)=\ha_3(\xi)= i \rx \sech(\rx \xi),\
\ha_2(\xi)=              \rx \tanh(\rx \xi),\
K_0'=\rx^4,\ K_1=0,\ 6 e_0=-\rx^2,
\label{eqFWM555xisolTrigoBright}
\end{eqnarray}
with $\rx,\xi_0,\varphi_0$ arbitrary.
In this eight-parameter solution,
$\rx$ is real,
and $\lambda_{12},\lambda_{34},\mu$ must be taken real
to enforce the complex conjugation between $A_j$ and $\barA_j$.

\textit{Remark 1}.
Despite the similarity with the stationary value
(\ref{eqFWM335_stationary_sol})
for this bright profile of the grating amplitude,
there is no limiting process yielding
(\ref{eqFWM335_stationary_sol}) from (\ref{eqFWM555xisolTrigoBright}).

\textit{Remark 2}.
For those solutions displaying constant phases for the amplitudes, 
there must exist a value of the damped sine-Gordon variable $u$,
Eq.~(\ref{eqdampedSG}),
able to represent the solution.
Up to the numerous additive and multiplicative constants
in (\ref{eqdampedSG}) and (\ref{eqFWM555xisol_Halphen}),
this value is essentially given by
\begin{eqnarray}
& &
e^{i u}=\ha_A(\xi),
\end{eqnarray}
in which $\ha_A$ and $\ha_\alpha$ are related by
the Landen transformation \cite[\S~16.14.2]{AbramowitzStegun}
\begin{eqnarray}
& &
\frac{\D}{\D \xi}\log \ha_A(\xi)
=-\frac{\ha_B(\xi) \ha_C(\xi)}{\ha_A(\xi)}=\ha_\alpha(\xi),
\end{eqnarray}
the correspondence between 
the elliptic invariants $(e_\alpha,e_\beta,e_\gamma)$ 
and $(e_A,e_B,e_C)$ being detailed in \cite[\S~16.14.1]{AbramowitzStegun}.
For the trigonometric degeneracy (\ref{eqFWM555xisolTrigoBright}),
the value is
\begin{eqnarray}
& &
e^{i u}=r \tanh r \xi,\
\end{eqnarray}
and the Landen transformation reduces to
the doubling of the argument with some shift,
\begin{equation}
\forall x:\
   \tanh x - \frac{1}{\tanh x}= -2 i \sech\left[2 x + i \frac{\pi}{2}\right],\
   \tanh x + \frac{1}{\tanh x}=  2   \tanh\left[2 x + i \frac{\pi}{2}\right].
\end{equation}

% ==================================================================
\section{Conclusion}
\label{sec:Conclusion}

The four-wave mixing has been characterized
by a lower-dimensional 
system of a deformed Maxwell-Bloch type.
Then the three and only three possibly singlevalued
limits of the four-wave mixing
model have been determined and integrated.
These consist of:
(i) the stationary case for any $\tau$ and $\gamma$;
(ii) the limiting case $1/\tau=0, \Re(\gamma)=0$
 which is identified to the complex unpumped Maxwell-Bloch system;
(iii) when $\Re(\gamma)=0$, the reduction $\xi=\sqrt{2 z} e^{-t / \tau}$
to an ODE system.
Those solutions which are localized (typically Jacobi bounded functions
$\sn$, $\cn$, $\dn$, $\cd$, $\nd$, $\sd$ \cite[\S~16.2]{AbramowitzStegun})
should improve both the design of the physical devices
to be manufactured
and the confidence in the numerical simulations.
As is often the case with methods based on singularities,
the present study cannot rule out possible closed form 
but multivalued solutions.

Moreover, the generic case $1/\tau\not=0, \Re(\gamma)\not=0$
has been shown to display a structure of singularities,
i.e.~of possible closed form solutions,
quite similar to that of the cubic complex Ginzburg-Landau equation.
These solutions will be investigated in a forthcoming paper. 

% ======================================================================
\section*{Acknowledgments}
We warmly acknowledge the financial support of the
Max-Planck-Institut f\"ur Physik komplexer Systeme,
where most of this work was performed,
and thank B.~Deconinck, V.Z.~Enol'skii and K.~Takemura for helpful advice.

% ************************************************************* References

% =============================================== APPENDIX
% ======================================================================
\section{Appendix. Complex amplitudes of the integrable $\xi$ reduction}
\label{sectionComplex_amplitudes}
\label{AppendixA}

By elimination from (\ref{eqFWMxi}),
both fields $\eer$ and $\ecr$ obey the same equation,
\begin{eqnarray}
\left(\frac{\D^2}{\D \xi^2} -2 (\wp(\xi)-e_0)\right) \psi=0,\ \psi=\eer, \ecr.
\label{eqLame1}
\end{eqnarray}
According to a classical result of Floquet, 
any linear differential equation with doubly periodic coefficients
admits at least one solution which is 
doubly periodic of the second kind \cite{HalphenTraite}.
The elementary unit of such doubly periodic functions of the second kind 
has been introduced by Hermite under the name 
\textit{\'el\'ement simple} 
$\Elemsimp(\xi,q,k)$ 
\cite[vol.~II, p.~506]{HalphenTraite},
\begin{eqnarray}
& &
\Elemsimp(\xi,q,k)=\frac{\sigma(\xi+q)}{\sigma(\xi) \sigma(q)} e^{(k-\zeta(q)) \xi},
\label{eqdefElemsimp}
\end{eqnarray}
chosen to have as only singularity a simple pole with residue $1$ at the origin.
Lam\'e indeed proved that Eq.~(\ref{eqLame1})
admits the two solutions
$\Elemsimp(\xi,-a,0)$ and $\Elemsimp(\xi,+a,0)$,
which are generically linearly independent.
Hence the complex amplitudes
\begin{eqnarray}
& & {\hskip -7.0 truemm}
\left\lbrace
\begin{array}{ll}
\displaystyle{
\eer= -i e^{ 2i \varphi_0} \Elemsimp(\xi,-a,0),\
\intImr= -\frac{i}{2 \invuniti} e^{ 2i \varphi_0} 
 \frac{\wp'(\xi)-\wp'(a)}{\wp(\xi)-\wp(a)} \Elemsimp(\xi,-a,0),\
}\\ \displaystyle{
}\\ \displaystyle{
\ecr= -i e^{-2i \varphi_0} \Elemsimp(\xi, a,0),\
\intIcr=-\frac{i}{2 \invuniti} e^{-2i \varphi_0} 
 \frac{\wp'(\xi)+\wp'(a)}{\wp(\xi)-\wp(a)} \Elemsimp(\xi, a,0),\
}
\end{array}
\right.
\label{eqFWM555xisolEIm}
\end{eqnarray}
in which 
the five constants of integration are
$e_0,g_2,g_3,\varphi_0$ and the origin of $\xi$.

Given the values (\ref{eqFWM555xisolEIm}) of $\eer(\xi),\ecr(\xi)$,
each variable $X,Y$ of the linear system (\ref{eqlinearXY})
also obeys a second order linear differential equation
with doubly periodic coefficients, e.g.,
\begin{eqnarray}
& &
\left(
\frac{\D^2}{\D \xi^2} 
- \frac{1}{2} \frac{\wp'(\xi)-\wp'(a)}{\wp(\xi)-\wp(a)} \frac{\D}{\D \xi}
- (\wp(\xi)-\wp(a))
\right) X=0,\ X=A_{1,{\rm r}}.
\label{eqodeX}
\end{eqnarray}
This equation has the same features as (\ref{eqLame1}):
unique singularity $\xi=0$ of the Fuchsian type,
Fuchs indices equal to $-1,1$,
absence of logarithms in the general solution.
A direct search for solutions of the elementary type (\ref{eqdefElemsimp})
provides the two solutions, generically linearly independent,
\begin{eqnarray}
& & {\hskip -13.0 truemm}
    X=\Elemsimp(\xi,+a/2 \pm \hh,0),\
    Y=\Elemsimp(\xi,-a/2 \pm \hh,0),\
 \wp    (\hh)=\wp(a/2)-2 \frac{{\wp'}^2(a/2)}{\wp''(a/2)}.
\label{eqFWM555xisolX}
\end{eqnarray}
Taking account of the first integrals
\begin{eqnarray}
    A_{1,{\rm r}} \barA_{1,{\rm r}} 
   +A_{2,{\rm r}} \barA_{2,{\rm r}}
    = \hbox{constant},\
    A_{3,{\rm r}} \barA_{3,{\rm r}} 
   +A_{4,{\rm r}} \barA_{4,{\rm r}}
    = \hbox{constant},\
\label{eqFWM555FirstIntegrals}
\end{eqnarray}
the general solution for the complex amplitudes
can be parametrized as
\begin{eqnarray}
& & {\hskip -7.0 truemm}
\left\lbrace
\begin{array}{ll}
\displaystyle{
    A_{1,{\rm r}}=i_0 
\left(\ \ a_{12}\Elemsimp(\xi,+a/2+\hh,0)
       \ +b_{12}\Elemsimp(\xi,+a/2-\hh,0)\right) e^{\displaystyle{+i \varphi_0}},\
}\\ \displaystyle{
    A_{2,{\rm r}}=i_0 
\left(\ \ a_{12}\Elemsimp(\xi,-a/2+\hh,0)
       \ +b_{12}\Elemsimp(\xi,-a/2-\hh,0)\right) e^{\displaystyle{-i \varphi_0}},\
}\\ \displaystyle{
\barA_{3,{\rm r}}=i_0 
\left(\ \ a_{43}\Elemsimp(\xi,+a/2+\hh,0)
       \ +b_{43}\Elemsimp(\xi,+a/2-\hh,0)\right) e^{\displaystyle{+i \varphi_0}},\
}\\ \displaystyle{
\barA_{4,{\rm r}}=i_0 
\left(\ \ a_{43}\Elemsimp(\xi,-a/2+\hh,0)
       \ +b_{43}\Elemsimp(\xi,-a/2-\hh,0)\right) e^{\displaystyle{-i \varphi_0}},\
}\\ \displaystyle{
\barA_{1,{\rm r}}=i_0 
\left(\ \ A_{12}\Elemsimp(\xi,-a/2-\hh,0)
         +B_{12}\Elemsimp(\xi,-a/2+\hh,0)\right) e^{\displaystyle{-i \varphi_0}},\
}\\ \displaystyle{
\barA_{2,{\rm r}}=i_0 
\left(   -A_{12}\Elemsimp(\xi,+a/2-\hh,0)
         -B_{12}\Elemsimp(\xi,+a/2+\hh,0)\right) e^{\displaystyle{+i \varphi_0}},\   
}\\ \displaystyle{
    A_{3,{\rm r}}=i_0 
\left(\ \ A_{34}\Elemsimp(\xi,-a/2-\hh,0)
         +B_{34}\Elemsimp(\xi,-a/2+\hh,0)\right) e^{\displaystyle{-i \varphi_0}},\
}\\ \displaystyle{
    A_{4,{\rm r}}=i_0 
\left(   -A_{43}\Elemsimp(\xi,+a/2-\hh,0)
         -B_{43}\Elemsimp(\xi,+a/2+\hh,0)\right) e^{\displaystyle{+i \varphi_0}},\
}\\ \displaystyle{
i_0^2=\frac{1}{\gnl\tau},
}
\end{array}
\right.
\label{eqFWM555xisolAj}
\end{eqnarray}
with
\begin{eqnarray}
& & {\hskip -7.0 truemm}
\left\lbrace
\begin{array}{ll}
\displaystyle{
a_{12}=\ \cos \mu \cosh \lambda_{12}\ e^{+i \alpha_{12}},\
A_{12}=\ \cos \mu \cosh \lambda_{12}\ e^{-i \alpha_{12}},\
}\\ \displaystyle{
b_{12}=i \cos \mu \sinh \lambda_{12}\ e^{+i  \beta_{12}},\
B_{12}=i \cos \mu \sinh \lambda_{12}\ e^{-i  \beta_{12}},\
}\\ \displaystyle{
a_{34}=\ \sin \mu \cosh \lambda_{34}\ e^{+i \alpha_{34}},\
A_{34}=\ \sin \mu \cosh \lambda_{34}\ e^{-i \alpha_{34}},\
}\\ \displaystyle{
b_{34}=i \sin \mu \sinh \lambda_{34}\ e^{+i  \beta_{34}},\
B_{34}=i \sin \mu \sinh \lambda_{34}\ e^{-i  \beta_{34}},\
}\\ \displaystyle{
e^{\displaystyle{i(\alpha_{12}-\beta_{12}-\alpha_{34}+\beta_{34})}}=\pm 1,\
\tan^2 \mu=\pm \frac{\sinh (2 \lambda_{12})}{\sinh (2 \lambda_{34})}.
}
\end{array}
\right.
\label{eqAj_constants}
\end{eqnarray}
The five additional integration constants are
three of the four constant real phases
$\alpha_{12}$, $\beta_{12}$, $\alpha_{34}$, $\beta_{34}$,
plus the two complex constants $\lambda_{12},\lambda_{34}$.
Finally,
the conditions that $\barA_{j,{\rm r}}$ be the complex conjugate of
$A_{j,{\rm r}}$ puts on $\lambda_{12},\lambda_{34}$
some constraints which depend on the choice for 
$\Elemsimp(\xi,\pm a/2\pm\hh,0)$, see text.

\vfill\eject

\end{document}